\documentclass[11pt]{article}

\usepackage{lmodern}
\usepackage[T1]{fontenc}
\usepackage{mathrsfs}
\usepackage{mathtools}
\usepackage{textcomp}
\usepackage{amsmath}
\usepackage{amsthm}
\usepackage{graphicx}
\usepackage{subfigure}
\usepackage{float}
\usepackage{amssymb}
\usepackage{physics}
\usepackage{color}
\usepackage{verbatim}
\usepackage{stackrel}
\usepackage[toc]{appendix}
\usepackage{vmargin}
\usepackage{xcolor}
\setpapersize{A4}
\setmargins{1.5cm}       
{0.5cm}                        
{18cm}                      
{26cm}                    
{0pt}                           
{1cm}                           
{0pt}                             
{1.5cm}                           
\usepackage[colorlinks=true, linkcolor=blue, citecolor=blue]{hyperref}
\usepackage{afterpage}
\newtheorem{theorem}{Result}
\newtheorem{definition}{Definition}

\date{}

\begin{document}
	\title{Quasi-equilibrium and quantum correlation in an open
		spin-pair system}
	\author{J.A. Taboada\footnote{agustintaboada8@gmail.com},
    H.H. Segnorile\footnote{segnorile@famaf.unc.edu.ar},
    C.E. González\footnote{ceciliae.gonzalez@unc.edu.ar},
    and R.C. Zamar\footnote{ricardo.zamar@unc.edu.ar}\\\\
    \emph{Facultad de Matemática, Astronomía, Física y Computación (FAMAF),}\\
    \emph{Universidad Nacional de Córdoba.}\\
    \emph{Instituto de Física Enrique Gaviola - CONICET.}\\
    \emph{\small (M. Allende y H. de la Torre - Ciudad Universitaria, X5016LAE - Córdoba, Argentina)}}
	\maketitle
	
\begin{abstract}
Quasi-equilibrium states that can be prepared in solids through Nuclear Magnetic Resonance (NMR) techniques are out-of-equilibrium states that slowly relax towards thermodynamic equilibrium with the lattice. In this work, we use the quantum discord dynamics as a witness of the quantum correlation in this kind of state. The studied system is a dipole interacting spin pair whose initial state is prepared with the NMR Jeener-Broekaert pulse sequence, starting from equilibrium at high temperature and high external magnetic field. It then evolves as an open quantum system within two different dynamic scenarios: adiabatic decoherence driven by the coupling of the pairs to a common phonon field, described within a non-markovian approach; and spin-lattice relaxation represented by a markovian master equation, and driven by thermal fluctuations. In this way, the studied model is endowed with the dynamics of a realistic solid sample.
The quantum discord rapidly increases during the preparation of the initial state, escalating several orders of magnitude compared with thermal equilibrium at room temperature. Despite the vanishing of coherences during decoherence, the quantum discord oscillates around this high value and undergoes a minor attenuation, holding the same order of magnitude as the initial state. Finally, the quantum discord dissipates within a time scale shorter than but comparable to spin-lattice relaxation.
\end{abstract}	
 

\section{Introduction}
	
	When an observed system is prepared in a non-equilibrium quantum state, a process immediately begins which eventually leads the system towards equilibrium with its environment. When that process involves a many-body system, the global evolution may proceed through a series of channels of different nature, acting on different time scales, and whose complexity may depend on the system, environment, and system-environment interactions.	
	A remarkable example is the class of quasi-equilibrium (QE) states that are readily prepared through solid-state Nuclear Magnetic Resonance (NMR) techniques \cite{Jeene_AdvMR68,Bonin_JCP2013}. Particularly, the well known Jeener-Broekaert pulse sequence allows preparing states characterized by one or a few constants of motion (or quasi-invariants). Such quasi-stationary states relax towards thermal equilibrium with characteristic times called relaxation times. Experiments show that three time scales can be distinguished in the evolution of QE states: preparation of an initial out-of-equilibrium condition, build-up of QE, and relaxation of the quasi-invariants. This work focuses to the quantumness of the successive states within these stages.
	
Observed NMR signals in the quasi-equilibrium state can be very precisely described by a density operator that is block diagonal in the eigenbasis of the spin system Hamiltonian. The phenomenology has been observed historically in solids \cite{Eisendrath_78,Keller_88} and more recently also in liquid crystals \cite{Bonin_JCP2013}. That is, on systems where the dipolar interaction determines the spin-spin dynamics.

The quantum correlations in QE states are still unknown. In particular, intriguing questions arise on the quantumness of macroscopically observable states that can be prepared through standard NMR techniques, and of the subsequent evolution. This is also matter of debate in other fields as the theory of equilibration \cite{RigolPRL_07,Nandkishore2016}.
In order to better understand the equilibration process and the nature of the quasi-stationary states, in this work we inquire about the quantumness of the transient states as they evolve to the final equilibrium, passing through the various stages of evolution. We analyze an assembly of nuclear spin pairs located at the sites of a crystal network, like the water protons in a solid hydrated salt. We use the quantum discord \cite{Olli_Zurek_01,Henderson_Vedral_2001} as a measure of the quantum correlation between the spins of the same pair along the whole time scale as the spin system evolves from an initial out-of-equilibrium quantum state towards thermal equilibrium.
This example illustrates the preparation of a quantum correlated state in a single crystal sample by means of the JB pulse sequence, and the subsequent evolution, subject to the action of quantum decoherence and thermal relaxation.

Quasi equilibrium states in solids and their relaxation dynamics is usually described in the framework of spin thermodynamics \cite{Jeene_AdvMR68,Keller_88,goldman1970spinT}, which is formally similar to a generalized grand canonical ensemble, where the state is written in terms of constants of motion or ``quasiinvariants''. However, this hypothesis does not focus on the mechanisms that lead an initial arbitrary state to QE.
In this work we use the  markovian master equation to describe the long term relaxation to thermal equilibrium. Instead, to describe the intermediate time scale dynamics (where the system attains QE), we adopt an alternative description based on quantum decoherence \cite{annals_21}, by extending to QE the conceptual framework recently developed to describe the irreversible attenuation of coherences in NMR time-reversal experiments.

The quantum discord dynamics of two-qubit systems  interacting with local reservoirs was addressed in \cite{TabeshManisc_QIP18,Maziero_Braz2013} and interacting with a common environment in \cite{Haikka_PRA13}. 
In an experimental study using NMR density matrix tomography techniques \cite{Maziero_Braz2013}, a Bell state is prepared on  spin pairs in liquids  where the J-coupling is the main spin-spin interaction and thermal noise due to the molecular environment motion acts as local reservoirs.
This type of system-environment coupling does not lead to intermediate QE states, due to the absence of a collective mechanism in liquids capable of removing coherence in the preferred basis.
Besides, the case of a spin 1/2 dipole coupled pair in presence of a magnetic field at thermal equilibrium was calculated as a function of temperature and the magnetic field \cite{Kuznetsova_QIP13}. The derivation was applied to a representative proton spin pair in gypsum and 1,2-dichloroethane, where the equilibrium quantum correlation is very small except for low temperatures.

In this work, we discuss the QD of dipole coupled spins in solids, where the spin dynamics is dictated by collective quantum decoherence (common environment) in the early time scale, and by spin-lattice relaxation (local thermal fluctuations) at longer times. In sections \ref{sec:model-evol} we describe the spin dynamics under these mechanisms. Section \ref{sec:QD} shows the calculation of the QD evolution and displays the results in the particular case of a gypsum single crystal. The relevance of the results is discussed in section \ref{sec:discussion}.

\section{Physical model: Hamiltonian and spin state evolution}
\label{sec:model-evol}

A realistic and at the same time mathematically manageable model  to approach quantum states prepared in solids, is that of dipole coupled spin 1/2 pairs system. The model mimics the single-crystalline sub lattice of hydration water protons occurring in hydrated salts like calcium sulfate dihydrate (gypsum), or potassium oxalate monohydrate (POMH), where the spin system can be considered an ensemble of weakly interacting spin pairs. We consider the system dynamics as that of an open quantum system along the various stages of its evolution. Thus, the total Hamiltonian has the general form
		\begin{equation}
		H = H_S + H_E + H_{SE},
		\label{ham total}
		\end{equation}
where $H_S$ represents the spin system, $H_E$ the environment, and $H_{SE}$ the system-environment coupling.
To model the spin dynamics within the different time scales, we adopt different formats for $H_{SE}$ and $H_E$, based on abundant NMR experimental information on hydrated crystals.
Since we are interested on quantum states that can be prepared in NMR experiments, let us assume that the main contributions to the spin system Hamiltonian are the Zeeman and the dipolar interactions
	\begin{equation}
	\label{Hs}
	H_S = H_Z + H_D^{\textrm{total}},
	\end{equation}
	where
$$ H_Z = \sum_{a} \omega_0 I_z^{(a)}$$
with $I_z^{(a)} = I_z^{(a1)}+I_z^{(a2)}$ the $z$-component of spin operator of the  $a$-th spin pair, $\omega_0 = \gamma B_0$ is the Larmor frequency, $B_0$ the static magnetic field, applied along the $z$ axis, and $\gamma$ the proton gyromagnetic ratio.
	Concerning the dipolar term, $H_D^{\textrm{total}}$, it is the sum of the intrapair and interpair contributions of the whole system
	\begin{equation} \label{H_D}
H_D^{\textrm{total}}=H_D+H_D^{\textrm{inter}},\quad H_D=\sum_{a} H_{D}^{(a)},\quad H_D^{\textrm{inter}}=\frac{1}{2}\sum_{a \neq b} H_{D}^{(ab)}
	\end{equation}
	We adopt the high magnetic field approximation $\lVert H_Z \lVert  \gg \lVert H_D \lVert$ and thus keep only the secular part of the dipole Hamiltonian $H_D^0$, which commutes with  $H_Z$. Thus,  
	\begin{equation}
	H_D^{(a)}= \sqrt 6 \;\Omega_0^{(a)}(\vec r_{12}^{\:(a)} ) \; T_{20}^{(a)},
	\label{HDA}
	\end{equation}
	with
	\begin{equation}
	\Omega_0^{(a)}(\vec r_{12}^{\:(a)} ) \equiv \frac{\mu_0\gamma^2\hbar}{8\pi  (r_{12}^{(a)})^3}\left[1- 3\cos^2\left(\theta_{12}^{(a)}\right)\right], \label{OmegaDA}
	\end{equation}
	where  $\theta_{12}^{(a)}$ is the angle between the external field and the internuclear vector $\vec r_{12}^{\hspace{0.5mm} (a)}$ between the interacting nuclei of pair $a$.
	$T_{20}^{(a)}$ is the zero component of the normalized irreducible spherical tensor of rank two 
	\begin{equation}
	T_{20}^{(a)} =\frac{2}{\sqrt{6}} \left[ 3 I_z^{a_1} I_z^{a_2} - {I}^{a_1}\cdot {I}^{a_2} \right],
	\label{T20}
	\end{equation}
	and the interpair terms are
	\begin{equation}
	H_{D}^{(ab)}= \sqrt{6}\sum_{uv}\Omega_0^{ab} (\vec r_{uv}^{\hspace{0.5mm}(ab)}) T_{20}^{(a_ub_v)},
	\label{HDAB}
	\end{equation}

The quantities $\Omega_0^{(a)} (\vec r_{12}^{\hspace{0.5mm} (a)} )$ y $\Omega_0^{(ab)} (\vec r_{uv}^{\hspace{0.5mm} (ab)})$ become scalars (the dipolar frequencies) when the spin system is either closed or is coupled to a stochastic environment, while should be considered as quantum operators when the open system is coupled to a quantum environment \cite{annals_21}.\\

In the following sections we discuss the dynamics of our model system along the whole process from preparation to thermal equilibrium. We write the time dependent density operators necessary to calculate the quantum discord. Throughout the different stages we consider the spins as a quantum system interacting  with its environment in a way which depends on the corresponding time scale. At the end of sec.\ref{sec:decoherence regime} we show that quantum decoherence drives the initial out-of-equilibrium state to a quasi-equilibrium state, and discuss the relationship with the spin thermodynamics viewpoint.

\subsection{Preparation}

This stage starts with the spin system in thermal equilibrium,  $\frac{1}{\cal Z}e^{- H_Z/kT} $, where $\cal Z$ is the partition function at temperature $T$, and proceeds with the Jeener-Broekaert pulse sequence (JB) \cite{PhysRev.157.232}, composed by a first pulse, free evolution during a short period $\tau$, and a second out-of-phase pulse.
Briefly, the first pulse creates single-spin single-quantum coherences in the spin system state, which evolve in the rotating frame mainly under the dipole spin-spin Hamiltonian during a time $\tau$. Along this period, multi-spin single-quantum coherences can develop, and the second pulse transforms part of the coherences just created into multi-spin order.	The fact that the time $\tau$ needed for preparing the state of interest is about 1/2 of the dipolar period, and thus, much shorter than the characteristic times of the subsequent irreversible processes (decoherence and relaxation), justify the following simplifying and not very restrictive assumptions: $i$) the spin system evolves as strictly isolated, that is, $H_{SE}=0$ within the preparation period; and $ii$) Its unitary evolution is mainly driven by the internal interaction $\sum_{a} H_D^{(a)}$ from \eqref{H_D}. In fact, in real systems of weakly interacting pairs the evolution of high-order multiple quantum coherences is very slow compared to the intra-pair dynamics \cite{Bonin_JCP2013}. Under these conditions, the formal analysis of the pairs system is equivalent to that of a single pair.
Thus, the initial equilibrium state of the spin pair, within the high temperature and high magnetic field approximation, is
	\begin{equation}
	\rho^{eq}
	\simeq \frac{1}{4}\left( \mathbb{I} +\beta_0I_z^{(a)} \right),
	\label{rhodequilib}
	\end{equation}
where $\beta_0 \equiv \frac{\hbar \omega_0}{k T}$, and $\omega_0 =\gamma B_0$ the Larmor frequency corresponding to an external field $ \vec{B}= B_0 \hat{z}$.
The first $(\pi/2)_y$ pulse rotates $I_z$ into $I_x$ and during the time $\tau$ the state evolves (in the rotating frame) under the dipolar Hamiltonian	
	\begin{equation}
	\rho(\tau)= \frac{1}{4}(\mathbb{I} + \beta_0 e^{-i\tau H_D^{(a)}}\:I_x^{(a)}\:e^{i\tau H_D^{(a)}} ),
	\end{equation}
which, in terms of spherical tensors $T_{LM}$ that involve the two spins of the $a$-th pair, is
	\begin{equation}
	\label{noastut}
	\rho(\tau)= \frac{1}{4}\left(\mathbb{I} - \beta_0  [-(T_{1,1}+ T_{1,-1})/ \sqrt{2} \cos \omega_D \tau + i(T_{2,1}+ T_{2,-1})\sin \omega_D \tau]\right),
	\end{equation}
where, in accordance with assumption ($i$),  $\omega_D=\Omega_0^{(a)}$ is the dipolar frequency. For the sake of concreteness, let the angle $\theta_{12}^{(a)}$ in \eqref{OmegaDA} be $\pi/2$, then, $\omega_D= \frac{3\,\mu_0\,\gamma^2\,\hbar}{16\,\pi\,r_{12}^3}$.

After the free evolution period, the second pulse rotates the state \eqref{noastut} by $\pi/4$ along the $x$ axis, so preparing the state
	\begin{equation}
	\label{rhodetau+}
	\rho(\tau^+)= \frac{1}{4}\left(1 - \beta_0  \left[ -(T_{1,1}+ T_{1,-1})/ \sqrt{2} \cos{\omega_D \tau} + \sin{\omega_D \tau}  (\sqrt{3/2} \;T_{2,0} + (T_{2,2}+ T_{2,-2})\right]\right).
	\end{equation}
It should be noticed that the state prepared with this sequence depends on the time $\tau$ and the pulse angle and phases.	Let us now consider $\rho(\tau^+)$ as the initial condition for the subsequent irreversible evolution.

\subsection{Decoherence regime} \label{sec:decoherence regime}	

During this intermediate scale, much longer than preparation but still earlier than Markovian relaxation, we abandon the picture of an isolated system and consider that the spin (pair) system is coupled to a quantum environment. This causes a non-unitary irreversible spin dynamics.
	It was recently shown \cite{annals_21,DomZamSegGzz17} that an open quantum system consisting of a partition of many equivalent elements with no direct interaction, in contact with a common bosonic bath can undergo a decoherent dynamics (in the adiabatic regime).
This full-quantum procedure revealed the occurrence of an irreversible process that explains qualitatively and quantitatively the observed behavior of the so-called ``NMR magic-echo'' reversal experiment. An important property of this energy-conserving decoherence mechanism is that it selectively preserves the density matrix elements laying on the diagonal in-blocks in the eigenbasis of the system-environment Hamiltonians (preferred basis), while attenuates the off-diagonal ones.  Accordingly, it is capable to bring an initial arbitrary state to a block diagonal form that does not evolve with the system Hamiltonian. In other words, it may thus explain the build-up of quasi-equilibrium states in a time scale earlier than thermal relaxation.

 The approach, in short, addresses the decoherent dynamics of an ensemble of pairs due exclusively to their contact with a common phonon
bath, and relies on the following assumptions: ($i$) The observed spin system is modeled by $H_S= \sum_a H_D^{(a)}$ (we neglect the interpair dipolar interactions from Eq.\eqref{H_D}),  and is coupled with a phonon bath, represented by
\begin{equation}
\label{eq:Ham_bos}
H_E ={\mathbb{I}}^{(s)}\otimes \sum_{k}
\omega_{k}\,{\bf b}_{k}^{\dagger}\;{\bf b}_{k},
\end{equation}
where  ${\bf b}_{k}$ are boson operators.
($ii$) There is no system-environment energy exchange (adiabatic regime) within this timescale, that is, $\left[H_S,H_{E} \right]=0$ and $\left[H_S,H_{SE} \right]=0 $, which implies that $\left\langle H_S\right\rangle $ is a constant of motion. The fact that  $\left[ H_E,H_{SE} \right] \neq 0$ imposes that the compound system-environment state evolution becomes not separable, and
evolves as a closed system. ($iii$)  The system-environment coupling  $H_{SE}$  accounts for the interaction of each observed system element with the common environment as
	\begin{equation} \label{H_SE}
	H_{SE} \equiv -\frac{3}{d}\sum_a H_D^{(a)}
	\otimes\sum_{k}\left((g_{k}^{(a)})^*{\bf b}_{k}+g_{k}^{(a)} {\bf b}_{k}^{\dagger}\right)\equiv \sum_a H_{SE}^{(a)},
	\end{equation}
where $d\equiv r_{12}^{(a)}$ is the distance between the spins within pair $a$. Notice that the interaction involves the sum over all the $a$ pairs and that the coupling coefficients $g_{k}^{(a)}$ encompass the collective phonon mode phases; this is
	\begin{equation*}
	\label{eq:def_gk_A_desc}
	g_{\bf k}^{(a)} = e^{-{\rm i}\vec{k}\cdot\vec{r}^{(a)}}g_{\bf k} ,
	\end{equation*}	
where $\vec{k}$ is the phonon wavenumber vector and the system-environment coupling strength $g_{\bf k}$ is independent of the pair position.\\
In this way, $H_{SE}$ represents the variation of the dipolar energy due to the atomic shifts related with collective excitations of the lattice.
It is worth emphasizing that this Hamiltonian is not written as a product of operators acting on each Hilbert subspace (spins and environment). Although each spin pair couples individually to the phonon modes (we do not consider the direct magnetic interaction between spins of different pairs), 
 the collective dynamic process governed by the lattice phonons establishes an indirect interaction that correlates pairs throughout the sample, which causes a non-unitary irreversible spin dynamics, in spite of the spin energy conservation.
	
The pair-phonon model outlined above, allowed writing the time dependent elements of the density matrix reduced to the $a$-th spin-pair as

	\begin{equation}
	\label{eq:Deltarho_boson-spin_aprox_fin}
		\rho^{(a)}_{m,n}(t) \simeq \rho^{(a)}_{m,n}(0)e^{2i\pi\omega_D\left(\kappa_{m} - \,\kappa_{n}\right)t}
	e^{-\left[\left(\kappa_{m} -\kappa_{n}\right)t/\tau_D\right]^2},
	\end{equation}
where $\{\left| m \right\rangle \}$ is the eigenbasis of the spin part of $H_{SE}^{(a)}$. In the case of the spin-environment coupling as \eqref{H_SE} coincides with the eigenbasis of the secular dipolar tensor
	\begin{equation}\label{eq:eigen_T20}
	{T}_{2,0}^{(a)}\left|m \right\rangle = \frac{1}{2\sqrt{6}}\;\kappa_{m}\,\left| m \right\rangle,
	\end{equation}
where the corresponding eigenvectors and eigenvalues are
\begin{equation}\label{eq:def_mA_kappaA}
\ket{m} = \left\{\begin{array}{l}
\ket{1,1} \equiv \ket{++}\\
\ket{1,0} \equiv \frac{\ket{+-}+\ket{-+}}{\sqrt{2}}\\
\ket{1,-1} \equiv \ket{--}\\
\ket{0,0} \equiv \frac{\ket{+-}-\ket{-+}}{\sqrt{2}}
\end{array}\right.\;
, \qquad \kappa_{m} = \left\{\begin{array}{c}
1 \\
-2 \\
1\\
0
\end{array}\right..
\end{equation}
Both the oscillation frequency and the attenuation time of each matrix element is determined by the eigenvalue difference $\kappa_{m} -\kappa_{n}$. The attenuation is also driven by a characteristic decay rate  $\tau_D^{-1}$ that depends on the system and environment physical constants as 		
	\begin{equation}
	\tau_D^{-1}= \frac{9\sqrt {2} }{16} \frac{c_s^2 m_p}{\omega_D^2\hbar\sigma},
	\end{equation}
with $c_s$ the speed of sound in the material of interest, $m_p$ the mass of the spin bearing nucleus, and $\sigma$ the eigenvalue distribution width.

Equation \eqref{eq:Deltarho_boson-spin_aprox_fin} implies that due to adiabatic decoherence the time evolution of a density matrix element is a damped oscillation, with a decay time that is proportional to the energy difference $ \kappa_{m}-\kappa_{n}$. This kind of selection based on the eigenvalues of the Hamiltonian that drives decoherence, or ``eigenselection'' \cite{SegZam11},  categorizes the eigenbasis of $H_{SE}$ as the preferred basis. The decay is purely due to the quantum correlated interaction between system and environment; formally, the coherence damping is associated with the relation $[H_{SE}, H_E] \ne0$ \cite{annals_21}.
 Physically, during the adiabatic decoherence regime there is no system-environment energy exchange; the system and its molecular environment interact quantum mechanically. 
The environment starts from thermal equilibrium, but its state does not remain static, it instead evolves due to a correlated interaction with the spins. The compound system-environment evolves in a correlated way as a whole, and all the spin physical magnitudes can be calculated in terms of the densiy matrix reduced over the environment. Further reduction to a single representative pair (over all other pairs) is necessary to calculate the quantum discord, however, it is worth to note that such double reduction keeps all the effects of the correlated spin-environment quantum interaction, which endows the simple pair-model with realistic solid-like characteristics.

The selective attenuation brings the initial state to one which is block diagonal in the preferred basis; it attenuates coherences in a characteristic time $\tau_D$ comparable with the time needed to attain quasi-equilibrium \cite{annals_21}.
In a general case, decoherence may bring the initial state into a combination of (traceless) operators ${\cal O}_j$ that do not evolve with $H_{SE}$ and are block diagonal in the preferred basis.  After decoherence, the density operator of Eq.\eqref{eq:Deltarho_boson-spin_aprox_fin} attains the general form
\begin{equation}\label{rho_QI}
\rho = \mathbb{I}- \sum_j \beta_j {\cal O}_j,
\end{equation}	
where $\beta_j$ are real coefficients  that derive from a purely quantum mechanical coupling of the spin system with a boson environment. The form of Eq. (\ref{rho_QI}) resembles the  density operator proposed in NMR in the context of the {\it spin temperature hypothesis}, where the $\beta_j$ are interpreted as inverse temperatures or generalized chemical potentials. The difference between Eq.\eqref{rho_QI} and the spin temperature view resides in the meaning of the coefficients, for the  decoherence derivation does not rely on thermodynamic arguments like maximizing the system entropy.
Since the diagonal terms are proportional to the dipolar Hamiltonian, and the preparation process amounts to amplifying them in comparison with equilibrium, it is traditionally said that the Zeeman order has been transferred to the dipolar order.

\subsection{Spin-lattice relaxation regime}
	\label{sec:Relaxation}

Within the longer time scale, the adiabatic condition is no longer
valid and the evolution is driven by thermal fluctuations. As usual in NMR, we adopt the Markovian relaxation theory to describe the long scale dynamics upon which the spin system reaches thermal equilibrium with the environment. During this regime, the spin system is in contact with a thermal reservoir. That is, the environment is assumed as a dissipative reservoir which is always at thermal equilibrium.

The density operator time dependence obeys a coarse-grain master equation which in the high temperature limit has the general form (in the interaction picture) \cite{abragam61},
\begin{equation}
\label{ecmaes}
\frac{d\rho(t)}{dt}= -\frac{1}{\hbar^2}\int_0^\infty dt' \hspace{0.5mm} \tr_E([H_{SE}^R(t),[H_{SE}^R(t'),\rho (t)]]\rho_E),
\end{equation}
where $\tr_E$ is the trace over the environment degrees of freedom and $\rho_E$ is the reservoir state at equilibrium.  The system-reservoir Hamiltonian $H_{SE}^R(t)$, relevant  within the relaxation time scale, accounts for the dipole-dipole interaction between spins belonging to different pairs. Commonly, the relaxation theory treats the spin degrees of freedom as quantum operators, and the lattice fluctuations due to thermal reorientation of the spin pairs as stochastic functions of the molecule orientation angles \cite{abragam61}.

Relaxation of the quasi invariants traditionally relies on the spin temperature hypothesis, which assumes that the system state can be described as a succession of quasi-equilibrium states. It was applied to the calculation of the Zeeman and dipolar relaxation times in gypsum and POMH \cite{Eisendrath_78,Keller_88,Look-Lowe}. The decoherence processes capable of bringing the spin system to a state of quasi-equilibrium in a short time scale with respect to the characteristic relaxation times provides a natural explanation for this hypothesis.

Experimental evidence on the dipolar relaxation in gypsum and POMH indicates that the fluctuating operator $H_{SE}^R(t)$ (thermal reorientation of hydration water molecules) induce singlet-triplet transitions in this kind of spin system. In order to account for this fact, the general form of $\rho(t)$ in Eq.\eqref{ecmaes} is
	\begin{equation}
\label{rhodiag}
\rho(t)=\frac{1}{\tr(\mathbb{I})}\left( \mathbb{I}-\beta_Z (t) H_Z -\beta_D(t) H_D-\beta_{N}(t) \Delta N_0\right),
\end{equation}
the JB sequence is tailored so that the coefficient $\beta_Z(0) =0$   \cite{JeenerBroekaert68};  $H_D$ was defined in Eq.\eqref{H_D}, and $\Delta N_0$
represents the deviation of the singlet occupation number from its thermal equilibrium value \cite{Keller_88}. The expectation value $ \left\langle \Delta N_0\right\rangle $ can be assumed zero at equilibrium and cannot be altered or manipulated by means of NMR pulses, however, molecular motions of an adequate symmetry may induce spin-lattice cross relaxation associated with the coupling between the triplet and singlet subspaces.

The problem discussed in this paper refers to the spin-lattice relaxation of the local quasi-invariants associated with the pairs: 
intra-pair dipole interaction energies, and excess population with respect to equilibrium of the singlet state.
Once the QE state is established, the relaxation mechanism pictured by $H_{SE}(t)$ induces cross relaxation of the intrapair $\beta_D$ and the $\beta_{N}$ coefficients via the thermal reservoir. By other hand, the Zeeman energy relaxes independently directly with the lattice.

Under these conditions, the description of spin-lattice relaxation proceeds: using \eqref{rhodiag} in \eqref{ecmaes} one gets two coupled equations for $\beta_D$ and $\beta_{N}$, which, in terms of dimensionless quantities

	\begin{equation}
	\label{defbetaD}
	x_D\equiv\beta_D\sqrt{\tr(H_D^2)/\tr(\mathbb{I})},
	\end{equation}
	\begin{equation}
	\label{defbetaN}
	x_N \equiv \beta_N \sqrt{\tr(\Delta N_0^2)/\tr(\mathbb{I})},
	\end{equation}
are
	\begin{equation}
	\label{ecacopD}
	\frac{d x_D}{d t}=-S_{DD}x_D-S_{DN}x_{N},
	\end{equation}
	\begin{equation}
	\label{ecacopN}
	\frac{d x_N}{d t}=-S_{NN}x_N-S_{DN}x_D,
	\end{equation}
where $S_{DD}$ and $S_{NN}$ coefficients represent the direct relaxation of each quasi-invariant with the reservoir, while $S_{DN}$ stands for the spin-lattice cross relaxation effect. The relaxation coefficients are \cite{Keller_88}
	\begin{equation}
	\label{SDD}
	S_{DD}= \frac{1}{\tr(H_D^2)}\int_0^\infty dt'\hspace{0.2mm}\tr([H_{SL}(t),H_D][H_{SL}(t-t'),H_D]\rho_E),
	\end{equation}
	\begin{equation}
	\label{SNN}
	S_{NN}=\frac{1}{\tr(\Delta N_0^2)}\int_0^\infty dt'\hspace{0.2mm} \tr([H_{SL}(t),\Delta N_0][H_{SL}(t-t'),\Delta N_0] \rho_E),
	\end{equation}
	\begin{equation}
	\label{SDN}
	S_{DN}= \frac{1}{\sqrt{\tr(\Delta N_0^2) \tr(H_D^2)}}\int_0^\infty dt'\hspace{0.2mm}\tr([H_{SL}(t),H_D][H_{SL}(t-t'),\Delta N_0]\rho_E).
	\end{equation}

\section{Quantum Discord Dynamics}\label{sec:QD}

Historically, the quantum discord (QD) is the first measure of quantum correlation beyond entanglement
\cite{Olli_Zurek_01,Henderson_Vedral_2001}. It is the result of extending the concept of conditional information to the quantum domain, and is defined as the difference between the quantum mutual information $I$ of a bipartite system $AB$ and the classical correlations $C(A|B)$
		\begin{equation}
		\label{eq:discord}
		Q(A|B)=I-C(A|B).
		\end{equation}
For the formal definition of QD we  follow \cite{Streltsov2015}, summarized in Appendix \ref{ap:definition}. Calculation, of QD involves optimization over all the Positive Operator Value Measure (POVM) acting on one of the system parts (say $B$). This is generally a difficult task, however,  POVMs reduce to projective measures for two-qubit states, and in the special case of $X$ states and Bell states, analytic expressions can be obtained \cite{PhysRevA.84.042313,PhysRevA.81.042105,PhysRevA.77.042303}.

In this section we use the time dependent density operator described in sec. \ref{sec:model-evol} to calculate the QD evolution along the \textbf{preparation}, \textbf{decoherence} and \textbf{relaxation} periods. Some details on the derivation of the following expressions are in Appendix \ref{ap:OptimalMeasurement}. In order to visualize the resulting time dependence, in section \ref{gypsum} we evaluate the QD in a concrete case.

\subsection{Preparation}
\label{sec:dipolar}

As pointed out in equation (\ref{rhodetau+}), the state of the system during the dynamics induced by the dipolar coupling is given by
\begin{equation}
\label{prep}
\rho(\tau^+)= \frac{1}{4}\left(\mathbb{I} - \beta_0  \left[ I_x(\cos{\omega_D \tau}) + \sin({\omega_D \tau})\left(\frac{1}{2}( T_{2,2}+ T_{2,-2})+\sqrt{\frac{3}{2}}T_{2,0}\right)\right]\right)
\end{equation}
We cast this density operator in terms of the Pauli matrices and apply the rotation sequence $R_z(\frac{\pi}{2})R_y(\frac{\pi}{2})$ to obtain
\begin{equation}
\label{eq:rhoU}
\begin{split}
	    \rho(\tau^+)&\equiv\dfrac{1}{4}\left(\mathbb{I}-\beta_0\left[\cos(\omega_D\tau)(\sigma_A^z+\sigma_B^z)+\sin(\omega_D\tau)\left(\dfrac{\sqrt{2}-1}{\sqrt{2}}\sigma_A^{z}\sigma_B^z-\dfrac{\sqrt{2}+1}{\sqrt{2}}\sigma_A^x\sigma_B^x+\sqrt{2}\sigma_A^y\sigma_B^y\right)\right]\right)\\
        &= \dfrac{1}{4}(\mathbb{I}+a_z\sigma_A^z+b_z\sigma_z^B+c_x\sigma_x^A\sigma_x^B+c_y\sigma_y^A\sigma_y^B+c_z\sigma_z^A\sigma_z^B)
	\end{split}
	\end{equation}
	This is a $X$ state with parameters
	\begin{equation}
	a_z=b_z=-\beta_0\cos(\omega_D\tau),\quad c_x=\frac{\sqrt{2}+1}{\sqrt{2}}\beta_0\sin(\omega_D\tau),\quad c_y=-\frac{\sqrt{2}-1}{\sqrt{2}}\beta_0\sin(\omega_D\tau),\quad c_z=-\sqrt{2}\beta_0\sin(\omega_D\tau)
	\end{equation}
	
	Appendix \ref{ap:OptimalMeasurement} shows that the optimal measurement for the QD is given by $\sigma_x$. Following \cite{PhysRevA.83.052108}, we find
	\begin{equation}
	\label{eq:min}
	\min_{\{\mathcal{M}_i^{(B)}\}}S(A|\{\mathcal{M}_i^{(B)}\})=h_+(r(t))+h_-(r(t))\approx 1-\frac{1}{2}r(t)^2
	\end{equation}
	where $h_{\pm}(r):=\frac{1}{2}-\frac{1}{2}(1\pm r)\log(1\pm r)$, $r:=\sqrt{c_x^2+a_z^2}$, and used that $r\ll1$, to keep terms up to second order in $r$. Using that $|a_z|\ll1$ gives
	\begin{equation}
	S(\rho_A)\approx 1-\frac{1}{2}a_z^2
	\end{equation}
	therefore, the classical correlations are
	\begin{equation}\label{CC_prep}
	C(\tau)=S(\rho_A)-\min_{\{\mathcal{M}_i^{(B)}\}}S(A|\{\mathcal{M}_i^{(B)}\})\approx 1-\dfrac{1}{2}a_z^2-1+\frac{1}{2}(c_x^2+a_z^2)=\dfrac{1}{2}c_x^2=\frac{3+2\sqrt{2}}{4}\beta_0^2\sin^{2}(\omega_D\tau)
	\end{equation}
	The eigenvalues of $\rho$ are
	\begin{equation}
	\label{eq:eigenvals}
	\lambda_{\pm}^{(1)}=\frac{1+c_z\pm\sqrt{4a_z^2+(c_x-c_y)^2}}{4}\equiv\frac{1+\Delta\lambda_{\pm}^{(1)}}{4},\quad\lambda_{\pm}^{(2)}=\frac{1-c_z\pm(c_x+c_y)}{4}\equiv\frac{1+\Delta\lambda_{\pm}^{(2)}}{4},\quad\Delta\lambda_{\pm}^{(j)}\ll1
	\end{equation}
	Using the approximation $\frac{1+x}{4}\log\frac{1+x}{4}\approx\frac{1}{4}\left(\frac{1}{2}x^2-x-2\right)$, it is easy to see that
	\begin{equation}
	\label{eq:S-I}
	S(\rho)=2-\sum_{j=1}^2\frac{(\Delta\lambda_{\pm}^{(j)})^2}{8}\Longrightarrow I(\tau)=\sum_{j=1}^2\frac{(\Delta\lambda_{\pm}^{(j)})^2}{8}-a_z^2
	\end{equation}
	Using equations (\ref{eq:min})-(\ref{eq:S-I}) in  (\ref{eq:discord}) allows writing the QD as a function of the preparation time as
	\begin{equation}\label{QD_prep}
	Q(\tau)=\sum_{j=1}^2\frac{(\Delta\lambda_{\pm}^{(j)})^2}{8}-a_z^2-\frac{1}{2}c_x^2=\sum_{j=1}^2\frac{(\Delta\lambda_{\pm}^{(j)})^2}{8}-\beta_0^2\left(\cos^2(\omega_D\tau)+\frac{3+2\sqrt{2}}{4}\sin^{2}(\omega_D\tau)\right)
	\end{equation}

\subsection{Decoherence}

The state evolves with time $t$ along this period as in Eq. \eqref{eq:Deltarho_boson-spin_aprox_fin}, which we write in a more convenient form as
\begin{equation}
\label{eq:deco}
\begin{split}
	\rho_{AB}(t)&=\dfrac{1}{4}\left[\mathbb{I}+\frac{\beta\omega_0\hbar}{4}\left\{2\cos(\omega_D\tau)e^{-9t^2/\tau_{X}^2}\cos(3\omega_Dt)(\sigma_A^{x}+\sigma_B^x)-e^{-9t^2/\tau_{X}^2}\cos(\omega_D\tau)\sin(3\omega_Dt)(\sigma_A^y\sigma_B^z+\sigma_A^z\sigma_B^y)\right.\right.\\
	&\left.\left.+\sin(\omega_D\tau)\left(\dfrac{\sqrt{2}-1}{\sqrt{2}}\sigma_A^{x}\sigma_B^x-\dfrac{\sqrt{2}+1}{\sqrt{2}}\sigma_A^y\sigma_B^y+\sqrt{2}\sigma_A^z\sigma_B^z\right)\right\} \right]
	\end{split}
	\end{equation}
	This is the result of applying (\ref{eq:Deltarho_boson-spin_aprox_fin}) to equation (\ref{prep}) and then writing it in terms of the Pauli matrices. Applying a rotation around the $x$-axis we take this state into the normal Bloch form, then performing the rotation sequence $R_z(\frac{\pi}{2})R_{y}(\frac{\pi}{2})$ gives
	\begin{equation}
	\label{eq:bloch-form-deco1}
	\rho_{AB}^{deco}(t)=\dfrac{1}{4}\left[\mathbb{I}+2\alpha(t)(\sigma_{A}^z+\sigma_{B}^z)+\sum_kc_k(t)\sigma_{A}^{(k)}\sigma_{B}^{(k)}\right],\quad k=\{x,y,z\}
	\end{equation}
	where
	\begin{equation}
	\alpha(t)=\beta_0e^{-9t^2/\tau_{X}^2}\cos(\omega_D\tau)\cos(3\omega_Dt)=:\alpha(t),\quad c_z=\beta_0\dfrac{\sqrt{2}-1}{\sqrt{2}}\sin(\omega_D\tau)
	\end{equation}
	\begin{equation}
    \label{eq:bloch-form-deco2}
	\begin{split}
	c_x(t)=-\beta_0\left(\dfrac{\sqrt{2}-1}{2^{3/2}}\sin(\omega_D\tau)+\sqrt{\dfrac{11+6\sqrt{2}}{8}\sin^2(\omega_D\tau)+e^{-18t^2/\tau_{X}^2}\cos^2(\omega_D\tau)\sin^2(3\omega_Dt)}\right)
	\end{split}
	\end{equation}
	\begin{equation}
    \label{eq:bloch-form-deco3}
	\begin{split}
	c_y(t)=-\beta_0\left(\dfrac{\sqrt{2}-1}{2^{3/2}}\sin(\omega_D\tau)-\sqrt{\dfrac{11+6\sqrt{2}}{8}\sin^2(\omega_D\tau)+e^{-18t^2/\tau_{X}^2}\cos^2(\omega_D\tau)\sin^2(3\omega_Dt)}\right)
	\end{split}
	\end{equation}
	As for the preparation period, the optimal measurement for QD is $\sigma_x$.
	
	The method for calculating the quantum discord is the same used in the previous stage, thus we just present the results.
	\begin{equation}\label{CC_deco}
	C(t,\tau)=\frac{1}{2}c_x^2(t)=\beta_0^2\left(\dfrac{\sqrt{2}+1}{2^{3/2}}\sin(\omega_D\tau)+\sqrt{\dfrac{11+6\sqrt{2}}{8}\sin^2(\omega_D\tau)+e^{-18t^2/\tau_{X}^2}\cos^2(\omega_D\tau)\sin^2(3\omega_Dt)}\right)^2
	\end{equation}
	The eigenvalues of state (\ref{eq:bloch-form-deco1}) are the same as those in equation (\ref{eq:eigenvals}), but now with the time dependence given by equations (\ref{eq:bloch-form-deco1})-(\ref{eq:bloch-form-deco2}). The mutual information and the QD are
	\begin{equation}\label{I_Q_deco}
	I(t,\tau)=\sum_{j=1}^2\frac{\Delta\lambda_{\pm}^{(j)}}{8}-\alpha(t,\tau)^2,\quad Q(t,\tau)=\sum_{j=1}^2\frac{\Delta\lambda_{\pm}^{(j)}}{8}-\alpha(t,\tau)^2-\frac{1}{2}c_x^2(t,\tau)
	\end{equation}

\subsection{Relaxation} \label{QD_relaxation}

Solving the equations of motion (\ref{ecacopD}) and (\ref{ecacopN}) reported in section \ref{sec:Relaxation}, the dynamic of the system is described by the following density operator
\begin{equation}
\label{eq:relax}
	\rho_{quasi}=\dfrac{1}{Z}\left(\mathbb{I}+\beta_0\left[\dfrac{1}{2}x_z(t)\sigma_{AB}^z+\dfrac{\sqrt{3}}{2}{x}_D(t)T_{20}-\dfrac{1}{\sqrt{3}}{x}_0(t)K_{00}\right]\right)
	\end{equation}
	where $\sigma_{AB}^z=\sigma_A^z+\sigma_B^z$, and $T_{20}=\dfrac{1}{2\sqrt{6}}(2\sigma_A^z\sigma_B^z-\Vec{\sigma}_A^{x,y}\cdot\Vec{\sigma}_B^{x,y})$, $K_{00}=-\frac{1}{2\sqrt{3}}\Vec{\sigma}_A\cdot\Vec{\sigma}_B$, and $Z=4$ is the partition function in the high temperature limit. The functions $x_z(t)$, $x_Q(t)$ and $x_0(t)$ are defined as follows
	\begin{equation}
	x_z(t)=1-e^{-t/T_z},\quad x_Q(t)=A_{L}e^{-t/T_L}+A_{S}e^{-t/T_S},\quad x_0(t)=\sqrt{A_{L}A_{S}}(e^{-t/T_L}-e^{-t/T_S})
	\end{equation}
	where the coefficients $A_{L,S}$ and relaxation times $T_{L,S}$ are purely temperature dependent. It is direct
	to see that $x_Z(t)\propto \langle H_z\rangle(t)$, $x_Q(t)\propto \langle H_D\rangle(t)$ and $x_0(t)\propto\langle\Delta N_0\rangle(t)$, e.i., the Zeeman energy, the dipolar energy and the excess of singlet. This is a very important and useful feature of quasi-equilibrium states as it allows us, as we will see later, to express QD in terms of these expectation values.
	
	We can cast (\ref{eq:relax}) into the \textit{X-state} form
	\begin{equation}
	\rho_{quasi}=\dfrac{1}{4}(\mathbb{I}+\beta_0'(a_z(t)(\sigma_A^z+\sigma_B^z)+c_z(t)\sigma_A^z\sigma_B^z+c_{xy}(t)\Vec{\sigma}_A^{x,y}\cdot\Vec{\sigma}_B^{x,y})),\quad\beta_0'=\frac{\beta_0}{2}
	\end{equation}
	where
	\begin{equation} \label{def_a_c}
	a_z(t):=2x_z(t),\quad c_z(t):=\frac{{x}_Q(t)}{\sqrt{2}}+x_0(t),\quad c_{xy}(t):={x}_0(t)-\dfrac{{x}_Q(t)}{2\sqrt{2}}
	\end{equation}
	Given that the coefficients $A_{L,S}$ as well as relaxation times $T_{L,S}$ depend on temperature in a non-trivial way, we cannot give an explicit expression for the QD without choosing a particular temperature $T$. To continue the calculation, we choose $T=200$ K and provide the corresponding expressions.
	
	
	Given that $c_x=c_y$, we have just one parameter to carry out the optimization, so it is convenient to do it by hand.  Following \cite{PhysRevA.83.052108}, the conditional entropy, after neglecting terms of order two and higher in $\beta_0$, is
	\begin{equation}
	\begin{split}
	S(A|\{\Pi_B\})\approx1-\dfrac{r_+^2+r_-^2}{4}&=1-c_{xy}(t)^2\sin(\theta)^2-\dfrac{1}{4}(a_z(t)+ c_z(t)\cos(\theta))^2-\dfrac{1}{4}(a_z(t)- c_z(t)\cos(\theta))^2\\
	&=1-c_{xy}(t)^2-\frac{1}{2}a_z(t)^2+\left(c_{xy}(t)^2-\frac{1}{2}c_z(t)^2\right)\cos(\theta)^2
	\end{split}
	\end{equation}
	where $r_{\pm}=\sqrt{c_{xy}^2(t)\sin^2(\theta)+(a_z(t)\pm c_z(t)\cos(\theta))^2}$.
	
	Because $a_z^2(t),c_{xy}^2(t)$ are both positive and $c_{xy}(t)^2-\frac{1}{2}c_z(t)^2<0$ for all $t$, we conclude that the optimal measurement for QD is given by $\sigma_z$. Then the classical correlations, mutual information and QD are
	\begin{equation}
	\label{eq:Qrelax}
	C(t)=\frac{1}{2}c_z^2(t),\quad I(t)=\sum_{j=1}^2\frac{(\Delta\lambda_{\pm}^{(j)})^2}{8}-a_z^2(t),\quad Q(t)=\sum_{j=1}^2\frac{(\Delta\lambda_{\pm}^{(j)})^2}{8}-a_z^2(t)-\frac{1}{2}c_{z}^2(t)
	\end{equation}
	
It is worth noticing that the classical correlations are quantified by $c_z^2(t)$ in the relaxation period (see Eq.\eqref{eq:Qrelax}) while they depend on  $c_x^2(t)$ in the previous stages (see Eqs.\eqref{CC_deco} and \eqref{CC_prep}) 	
This is a direct consequence of being $\sigma_z$ the optimal measurement for QD and not $\sigma_x$. This discontinuity in the optimal measure between the decoherence and the relaxation regimes is a consequence of disregarding the inter-pair interaction within the decoherence model, which, in turn, preserves double-quantum coherences as if they were true quasi-invariants. This makes state \eqref{eq:relax} to be different from the asymptotic state in \eqref{eq:deco}.

\subsection{Example}\label{gypsum}

We now evaluate the QD using the parameters corresponding to a gypsum single crystal oriented with respect to the magnetic field so that all the pairs are magnetically equivalent. The parameters used are: $B=20$MHz, $T=200$K, $\omega_D=95.5$kHz.
We use the experimental decoherence rate $\tau_D=306\mu$s \cite{annals_21}. The experimental relaxation times at this temperature and magnetic field strength are $T_L=12$ms, $T_s=0.6$ms, $A_L=0.88$ $A_s=0.12$ and $T_z=512$ms \cite{Eisendrath_78,footnote_TLTS}.

\begin{figure}[htp]
\centering
\includegraphics[width=.32\textwidth]{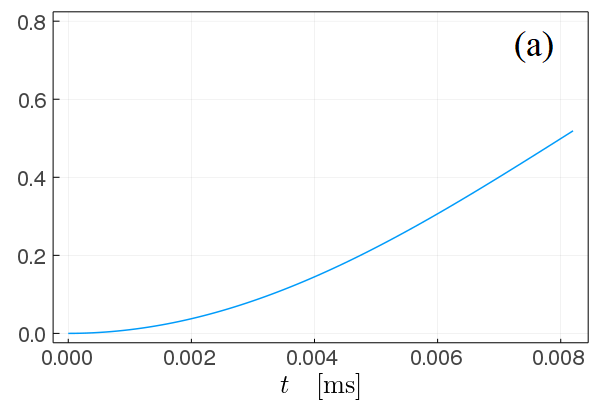}\hfill
\includegraphics[width=.32\textwidth]{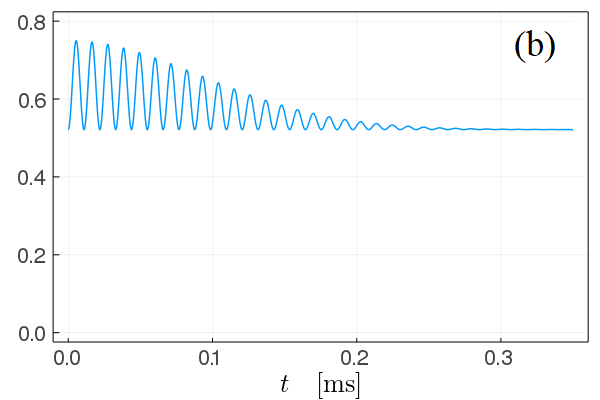}\hfill
\includegraphics[width=.32\textwidth]{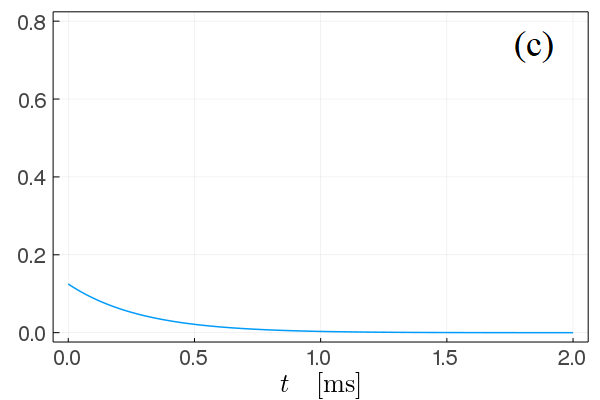}
\caption{Quantum discord dynamics in units of $\beta_0^2$ within each stage of the evolution.}
\label{fig:Q}
\end{figure}

Figure \ref{fig:Q} shows the QD  in units of $\beta_0^2$ as a function of convenient dimensionless time parameters along the three stages of evolution.

	We see a marked difference between each scenario. In fig.  \ref{fig:Q}(a) The preparation time is chosen as $\tau=\pi/4\omega_D$ and QD comes from Eq.\eqref{QD_prep}.
 Mostly all the correlation is built up during preparation, and decays to zero during relaxation, given by Eq.\eqref{eq:Qrelax}. Within the decoherence regime, in  fig. \ref{fig:Q}(b) the QD  undergoes an attenuated oscillation and reaches a steady value when attaining the quasi-equilibrium state, according to Eq.\eqref{I_Q_deco}.

It is worth noticing that preparation brings the QD  from its equilibrium value to a magnitude of the order of $\beta_0^2 \sim 4.8\times 10^{-12}$. According to ref. \cite{Kuznetsova_QIP13} the equilibrium value of QD in each spin pair at $T \sim 200K $ is as small as QD$_{eq}\sim  4.5\times10^{-18}$. Contrastingly, the QD at the end of the preparation period at the same temperature and in the standard magnetic field used in our calculation is QD $\sim 10^{-12}$. That is, the sole evolution under the JB pulse sequence during the time $\tau$ augments the quantum discord by six orders of magnitude.

\begin{figure}[htp]
	\centering
	\includegraphics[width=.5\textwidth]{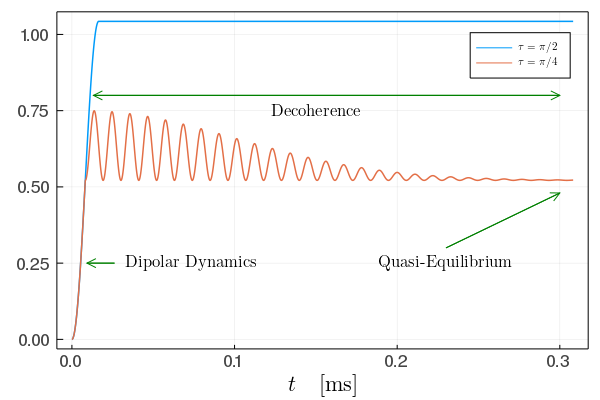}
	\caption{Quantum discord in units of $\beta_0^2$, for two different preparation times: $\tau=\pi/4\omega_D$ (red), and $\tau=\pi/2\omega_D$ (blue).}
	\label{fig:sequences}
\end{figure}

During decoherence, the QD undergoes a damped oscillation with a frequency $\sim \omega_D$. It rises above the initial value and then decays to the same value marked with a horizontal dashed line in fig. \ref{fig:Q}. The oscillation comes from the contribution of off-diagonal elements of the density matrix (in the preferred basis), while the constant value is related to the block diagonal contribution, which is insensible to decoherence. A convenient way to prepare an initial free-from-decoherence state is to use an optimal preparation time $\tau=\pi/2\omega_D$, which is the one that also yields the maximum dipolar energy expectation value.
In such a manner, one gets the maximum value for QD. Figure \ref{fig:sequences} compares the evolution within preparation and decoherence for two settings of $\tau$. The red line rises to a higher value during preparation, and more importantly, it stays unchanged during decoherence.

	\begin{figure}[h!]
	\centering
	\includegraphics[width=.5\textwidth]{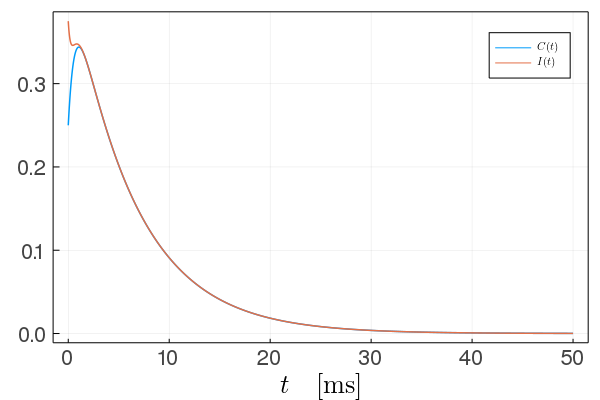}
	\caption{Mutual information (blue) and Classical Correlations (red)  in units of $\beta_0^2$, as a function of time within the relaxation regime.}
	\label{fig:I-C}
\end{figure}

Figure \ref{fig:I-C} shows the Mutual information $I(t)$  and Classical Correlations $C(t)$ in units of $\beta_0^2$, as a function of time within the relaxation regime. Unlike the QD, they do not show a monotonic decay. Even during this  purely dissipative process some classical correlations arise in the early relaxation regime, and decay later. The mutual information also shows a similar behavior but in a less sharp way.
Notice also that the characteristic decay time of $I(t)$ and $C(t)$ is much longer than that of the QD.

\begin{figure}[H]
		\centering
		\includegraphics[width=.5\textwidth]{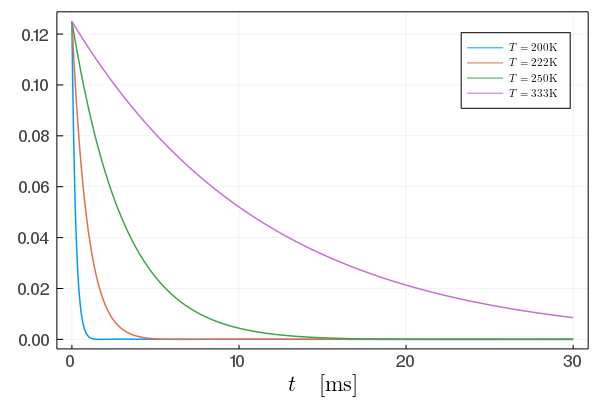}
		\caption{QD as a function of time and temperature  (each curve is normalized to the respective $\beta_0$).}
		\label{fig:Q(T)}
	\end{figure}

Let us now analyze the dependence of QD on temperature during relaxation (the earlier stages do not depend on $T$).
The QD decay time depends on the relative value of $T_L$ and $T_S$. We use the relaxation times temperature dependence reported in the classic work of ref. \cite{Eisendrath_78} to calculate the QD curves in fig.\ref{fig:Q(T)}. Accordingly, within the range $200\textrm{K}\leq T \leq300$K,  longer QD attenuation times correspond to higher temperatures. Though disputing with intuition, this indicates that cooling down the system makes relaxation more efficient and correlations more fragile. Of course this trend is valid within a rather high temperature regime.

\section{Discussion} \label{sec:discussion}
	
We used the quantum discord evolution as a witness of the quantum correlation of states prepared on a system of spin pairs through successive stages: preparation with an NMR pulse sequence, starting from equilibrium at high temperature and external magnetic field; decoherence regime, where the reduced-matrix elements dynamics is driven by the coupling of the pairs to a common phonon field; and relaxation regime, driven by thermal fluctuations (spin-pair flips).
In this way, the studied spin pair undergoes the dynamics of a realistic solid. The calculation of QD involves structure parameters, relaxation times and decoherence times corresponding to an actual gypsum single crystal.

The quantum discord has a dramatic increase during preparation of the initial state, escalating several orders of magnitude compared with that of the thermal equilibrium at room temperature. Since maximum QD achievable through the JB sequence is of course limited by the initial equilibrium Zeeman energy, that is, by $\beta_0$, a moderate improvement may be accomplished by increasing the external field strength, or by increasing the equilibrium polarization by other means. During decoherence it
oscillates around this high value and undergoes a minor attenuation, holding the same order of magnitude as the initial state. Once the quasi-equilibrium state is established, the quantum discord finally decays within the time scale of spin-lattice relaxation. However, if the QD lifetime is concerned,  one should also take into consideration that the relaxation times rest on the spectral densities of thermal fluctuations of the reservoir and that the resulting temperature dependence of relaxation times and thus, the QD, reflects such relationship (see Fig. \ref{fig:Q(T)}). As shown in section \ref{QD_relaxation},  also the occurrence of spin-lattice cross relaxation mechanisms strongly limits the lifetime of QD.

Eq. \ref{eq:Qrelax} provides a way of measuring QD experimentally, as well as $I(t)$ and $C(t)$. The three quantities depend on $c_z(t)$ and $a_z(t)$, which in turn are functions of the observable amplitudes $A_Z$, $A_L$ and $A_S$ (through $x_Z, x_D$ and $x_0$, see Eq. \eqref{def_a_c}). For example, $I(t)$ depends on the Zeeman energy, which can be measured by a traditional inversion experiment \cite{Levitt:500323}. Besides, the Jeener-Brokaert sequence described above was precisely designed to observe the dipolar relaxation, and provides the parameters $A_L, A_S, T_L$ and $T_S$ \cite{Eisendrath_78}. In this way,  the quasi-equilibrium state has the advantage that might be experimentally determined through standard NMR procedures with no need of employing more demanding experiments like the tomography of the density matrix.

The model system of non-interacting spin pairs considered here entails the fact that the double quantum coherence term $(T_{2,2}+ T_{2,-2})$ becomes a constant of the motion. This characteristic is not consistent with experimental observation and can be circumvented by including some inter-pair interaction terms both in the spin Hamiltonian $H_S$ and in the spin-environment Hamiltonian $H_{SE}$ (a generalization of the decoherence mechanism which includes terms $I_z^A I_z^B$ will be published elsewhere). {Nevertheless, this fact opens the question of whether or not quasi-invariants are good states to protect quantum correlations from decoherence in a general setting, i.e., under a broader dynamic. This question is a matter of quantum control which of course it cannot be answered beforehand as it depends on the available control techniques and on the internal interactions. }

It is worthy of comment that despite the apparent simplification that entails keeping just one pair from the whole ensemble,  such a pair carries information about the entire ensemble [10]. Relaxation is also a collective process that involves the whole ensemble's presence as seen from the expressions for the relaxation times. In a nutshell, the results presented here show how the dynamic of the QD is affected not just by being coupled to an environment but also by being a small part of a larger system.

In order to appraise the magnitude of the QD calculated in figures \ref{fig:sequences} and \ref{fig:Q(T)}, let us give an estimate of the intra-pair dipolar energy expectation value  $\left\langle H_D^0\right\rangle_{QE}$ in the QE state and compare it with that of the final equilibrium $\langle H_D^0\rangle_{equil}$. Assuming that we use the optimal preparation time and that decoherence left the state in the invariant form
\begin{equation}
\label{astuteq}
\rho_{QE}= \frac{1}{\tr(\mathbb{I})}\left( \mathbb{I}+\beta_R\omega_0 \sqrt{\frac{3}{2}}T_{2,0}\right)  = \frac{1}{\tr(\mathbb{I})} \left(\mathbb{I}+\frac{3}{2}\beta_R \frac{\omega_0}{\omega_D} H_D^0\right),
\end{equation}
where we used that $H_D^0 =\sqrt{2/3}\;\omega_D T_{20}$. Then, the quotient of expectation values is proportional to the ratio of the Larmor and dipolar frequencies
\begin{equation}
\label{valdexpect}
\frac{\langle H_D^0\rangle_{qe}}{\langle H_D^0\rangle_{equil}}=\frac{3 }{2}\frac{\omega_0}{\omega_D} \sim 10^3.
\end{equation}
This means that in a sample like gypsum the process of preparation of QE increased the dipolar energy expectation value in three orders of magnitude.
In the language of spin thermodynamics, this fact is interpreted as having lowered the ``dipolar spin temperature'' $T_D$ to roughly $T_D \equiv \frac{2}{3}\frac{\omega_0}{\omega_D}T \sim 0.2K$. According to reference \cite{Kuznetsova_QIP13} the quantum discord at equilibrium, in the high-temperature limit
(which, in gypsum corresponds to $T> 10 \mu K$ ), depends as ${\cal Q} \sim \frac{D^2}{4 \ln 2}(kT)^{-2}$. By mapping such expression to the water molecule in gypsum at a thermodynamic temperature $T=T_D$ yields a value  with a similar magnitude to the QD calculated in our figs. \ref{fig:sequences} and \ref{fig:Q(T)}, that is, ${\cal Q}(T=T_S) \sim Q $. This shows how efficient the JB sequence is to prepare a state with an amount of quantum discord comparable to a significantly low sample temperature.

In summary, we study the quantum discord dynamics through the equilibration process of a spin system in solids in presence of a magnetic field. We described the dynamics by considering the spins as an open quantum system, and using a realistic model along preparation of an initial state, early decoherence and spin-lattice relaxation towards thermal equilibrium. As an important result, we find that the quantum discord of the evolving state attains much greater values in comparison with the initial thermal equilibrium, and that the quantum correlation remains during the decoherence regime. The quasi-equilibrium quantumness finally dissipates due to the action of thermal fluctuations, with a greater rate than the relaxation of the dipolar energy expectation value.
The origin of getting high values of QD roots in having prepared single-quantum multi-spin coherences during the evolution after the first pulse of the JB sequence, and transforming them into multiple quantum coherences and quasi-invariants with the second pulse. After that, decoherence attenuates the non-diagonal terms of the density matrix in the preferred basis and keeps the diagonal part intact, so establishing a quasi-equilibrium state. This notable characteristic of the adiabatic decoherence process, associated with eigenselection, is a consequence of having a system-environment coupling as the one of Eq.\eqref{H_SE} which is essentially different from quantum noise models (that may hold for liquid systems). Despite the vanishing of coherences, the QD holds a markedly augmented value with respect to equilibrium, during  the whole decoherence regime.
 The dipolar network complexity and the preparation time ultimately dictate the characteristics of the QE state that is obtained with this procedure. In this work we show that they also determine the state quantum correlation.
It is worth to mention that the strategy (model system, dynamics and calculation of the QD) used throughout this work is directly generalizable to more complex systems, such as liquid crystals.

\section{Acknowledgement}

This work was supported by SECYT, Universidad Nacional de Córdoba. J.A.T. and  H.H.S. thank CONICET for financial support.

\appendix

\section{Definition of quantum discord} \label{ap:definition}

\begin{definition}
For a state $\rho$ we define the \textit{von Neumann entropy} as
\begin{equation}
S(\rho):=-\tr(\rho\log\rho)=-\sum_{i}\lambda_i\log\lambda_i
\end{equation}
where $\{\lambda_i\}$ are the eigenvalues of $\rho$.
\end{definition}

\begin{definition}
	Let us consider a bipartite system $AB$ described by a density operator $\rho_{AB}$. The \textit{mutual information} is defined as follows
	\begin{equation}
	\label{info_mutua}
	I:=S(\rho_A)+S(\rho_B)-S(\rho_{AB});\quad S(\rho_{AB})=-\tr(\rho_{AB}\log\rho_{AB}),\quad \rho_{A,B}=\tr_{B,A}\rho_{AB}
	\end{equation}
	where ``$\tr_{B,A}$'' stands for partial trace over the variables of system $B$ or $A$ respectively.
\end{definition}
The quantity (\ref{info_mutua}) represents the amount of the total correlations between systems $A$ and $B$.

\begin{definition}
	\label{CI}
	Given a bipartite system $AB$ and a POVM\footnote{POVM: Positive Operator Value Measure.} $\{\mathcal{M}_i^{(B)}\}$ acting on subsystem $B$, we define the \textit{conditional information} or \textit{conditional entropy} as
	\begin{equation}
	S(A|\{\mathcal{M}_i^{(B)}\}):=S(\rho_{A|\mathcal{M}_i^{(B)}})
	\end{equation}
\end{definition}
Finally, we follow with the next definition.

\begin{definition}
	\label{def IG}
	Given a bipartite system $AB$ we define the ``\textit{information gained}'' about $A$, given that the POVM $\{\mathcal{M}_i^{(B)}\}$ was performed on $B$, as
	\begin{equation}
	J(A|B)_{\{\mathcal{M}_i^{(B)}\}}:=S(\rho_A)-S(A|\{\mathcal{M}_i^{(B)}\})
	\end{equation}
\end{definition}

\begin{definition}
	Consider a bipartite system $AB$, the Quantum Discord is defined as:
	\begin{equation}
	\label{QD}
	Q(A|B)=I-\max_{\{\mathcal{M}_i^{(B)}\}}J(A|B)_{\{\mathcal{M}_i^{(B)}\}}=S(\rho_B)-S(\rho_{AB})+\min_{\{\mathcal{M}_i^{(B)}\}}S(A|\{\mathcal{M}_i^{(B)}\})
	\end{equation}
\end{definition}

We have that $C(A|B):=\max_{\{\mathcal{M}_i^{(B)}\}}J(A|B)_{\{\mathcal{M}_i^{(B)}\}}$ satisfies all these conditions. The general agreement is that $C(A|B)$ can be thought of as a measure of the classical correlations between $A$ and $B$ and consequently, we can write (\ref{QD}) as
\begin{equation}
Q(A|B)=I-C(A|B)
\end{equation}

\section{Determination of the Optimal Measurement}
\label{ap:OptimalMeasurement}

In this appendix, we give the missing calculation concerning QD and in particular those involving the determination of the optimal measurement for the preparation and decoherence stage. For this, we use the following result

\begin{theorem}
	\label{theo1}
	The optimal measurement for the QD of a real X state $\rho_X$ such that $|c_x|\ge|c_y|$ is
	
	(i) $\sigma_A^z$ if:
	\begin{equation}
	\left(\frac{|c_x+c_y|+|c_x-c_y|}{4}\right)^2\le\frac{c_z^2-b_z^2}{4}
	\end{equation}
	\hspace{5mm}(ii) $\sigma_x$ if:
	\begin{equation}
	\bigg|\sqrt{(1+c_z)^2-(a_z+b_z)^2}-\sqrt{(a_z-b_z-c_z+1)(b_z-a_z-c_z+1)}\bigg|\le|c_x+c_y|+|c_x-c_y|
	\end{equation}
\end{theorem}

\subsection{Dipolar Evolution}

Recall section \ref{sec:dipolar} and eq. \ref{eq:rhoU}, the state of the system is
\begin{equation}
\rho(\tau^+)=\dfrac{1}{4}\left(I+a_z\sigma_A^z+b_z\sigma_B^z+\sum_ic_i\sigma_A^i\otimes \sigma_B^i\right)
\end{equation}
where
\begin{equation}
a_z=b_z=-\beta_0\cos(\omega_D\tau),\quad c_x=\frac{\sqrt{2}+1}{\sqrt{2}}\beta_0\sin(\omega_D\tau),\quad c_y=-\frac{\sqrt{2}-1}{\sqrt{2}}\beta_0\sin(\omega_D\tau),\quad c_z=-\sqrt{2}\beta_0\sin(\omega_D\tau)
\end{equation}
Condition (i) reads as
\begin{equation}
\frac{3+2\sqrt{2}}{8}\sin^2(\omega_D\tau)\le\frac{1}{4}\left[\left(\frac{3}{2}-\sqrt{2}\right)\sin^2(\omega_D\tau)-\cos^2(\omega_D\tau)\right]
\end{equation}
and it is easy to see that this does not hold for the time interval we are interested in. Condition (ii), on the other hand, needs a bit of calculation
\begin{equation}
\begin{split}
\sqrt{(1+c_z)^2-(a_z+b_z)^2}&=\sqrt{\left(1+\frac{\sqrt{2}-1}{\sqrt{2}}\beta_0\sin(\omega_D\tau)\right)^2-2\beta_0^2\cos^2(\omega_D\tau)}\\
&=\sqrt{1+\frac{\sqrt{2}-1}{\sqrt{2}}2\beta_0\sin(\omega_D\tau)+\left(\frac{\sqrt{2}-1}{\sqrt{2}}\beta_0\sin(\omega_D\tau)\right)^2-2\beta_0^2\cos^2(\omega_D\tau)}\\
&\approx \sqrt{1+\frac{\sqrt{2}-1}{\sqrt{2}}2\beta_0\sin(\omega_D\tau)}\approx 1+\frac{\sqrt{2}-1}{\sqrt{2}}\beta_0\sin(\omega_D\tau)=1+c_z
\end{split}
\end{equation}
where we used that $\beta_0\ll1$ to discard those terms of order two and higher in $\beta_0$.
\begin{equation}
\bigg|\sqrt{(1+c_z)^2-(a_z+b_z)^2}-\sqrt{(a_z-b_z-c_z+1)(b_z-a_z-c_z+1)}\bigg|\approx2|c_z|=\frac{2(1-\sqrt{2})}{\sqrt{2}}|\beta_0\sin(\omega_D\tau)|
\end{equation}
\begin{equation}
\label{eq:condition2}
\Longrightarrow 2|c_z|\le\frac{2(1+\sqrt{2})}{\sqrt{2}}\beta_0|\sin(\omega_D\tau)|=|c_x+c_y|+|c_x-c_y|
\end{equation}
Thus, the optimal measurement for the QD is given by $\sigma_x$.

\subsection{Decoherence}

For decoherence we have
\begin{equation}
\label{eq:bloch-form-deco}
\rho_{AB}^{deco}(t)=\dfrac{1}{4}\left[I+2\alpha(t)(\sigma_{A}^z+\sigma_{B}^z)+\sum_k\lambda_k(t)\sigma_{A}^{(k)}\sigma_{B}^{(k)}\right]
\end{equation}
where we defined
\begin{equation}
a_z(t)=b_z(t)=\beta_0e^{-9t^2/\tau_{X}^2}\cos(\omega_D\tau)\cos(3\omega_Dt)=:\alpha(t),\quad c_z=\beta_0\dfrac{\sqrt{2}-1}{\sqrt{2}}\sin(\omega_D\tau)
\end{equation}
\begin{equation}
\begin{split}
c_x=-\beta_0\left(\dfrac{\sqrt{2}-1}{2^{3/2}}\sin(\omega_D\tau)+\sqrt{\dfrac{11+6\sqrt{2}}{8}\sin^2(\omega_D\tau)+e^{-18t^2/\tau_{X}^2}\cos^2(\omega_D\tau)\sin^2(3\omega_Dt)}\right)
\end{split}
\end{equation}
\begin{equation}
\begin{split}
c_y=-\beta_0\left(\dfrac{\sqrt{2}-1}{2^{3/2}}\sin(\omega_D\tau)-\sqrt{\dfrac{11+6\sqrt{2}}{8}\sin^2(\omega_D\tau)+e^{-18t^2/\tau_{X}^2}\cos^2(\omega_D\tau)\sin^2(3\omega_Dt)}\right)
\end{split}
\end{equation}
Now let us see if whether $\sigma_z$ or $\sigma_x$ is the optimal measurement for QD.

Condition (i) read as
\begin{equation}
\label{eq:condition}
\begin{split}
Izq:&=\frac{1}{4}\left(\frac{\sqrt{2}+1}{\sqrt{2}}|\sin(\omega_D\tau)|+\sqrt{\frac{11+6\sqrt{2}}{8}\sin^2(\omega_D\tau)+e^{-18t^2/\tau_X^2}\cos^2(\omega_D\tau)\sin^2(3\omega_Dt)}\right)^2\\
&\le\frac{1}{8}\left(\frac{(\sqrt{2}-1)^2}{\sqrt{2}}|\sin(\omega_D\tau)|-e^{-18t^2/\tau_X^2}\cos^2(\omega_D\tau)\cos^2(3\omega_Dt)\right)=:Der
\end{split}
\end{equation}
e.i., $Izq\le Der$.
\begin{figure}[h!]
	\centering
	\includegraphics[width=9cm]{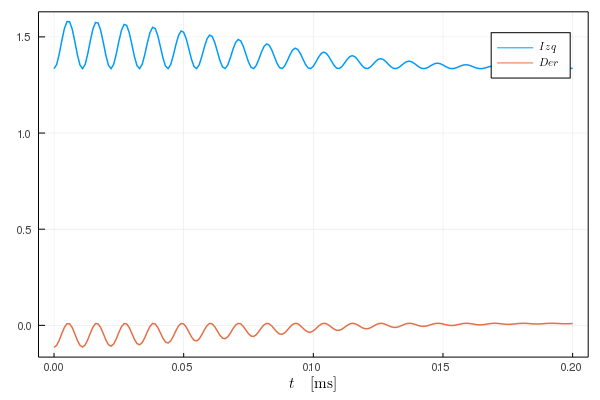}
	\caption{Time dependence of functions $Izq(t)$ and $Der(t)$ in the case $\omega_D\tau=\frac{\pi}{4}$.}
	\label{fig. 1}
\end{figure}

Figure \ref{fig. 1} shows that inequality  \ref{eq:condition} does not hold at all. To check condition (ii), we follow the same procedure as before, we keep just the linear term in $\beta_0$. Then, the condition reduces again to
\begin{equation}
2|c_z|\le |c_x+c_y|+|c_x-c_y|
\end{equation}
\begin{equation}
\begin{split}
|c_z|&=\frac{\sqrt{2}-1}{\sqrt{2}}|\sin(\omega_D\tau)|\\
&\le\left(\frac{\sqrt{2}+1}{\sqrt{2}}|\sin(\omega_D\tau)|+\sqrt{\frac{11+6\sqrt{2}}{8}\sin^2(\omega_D\tau)+e^{-18t^2/\tau_X^2}\cos^2(\omega_D\tau)\sin^2(3\omega_Dt)}\right)\\
&=\frac{|c_x+c_y|+|c_x-c_y|}{2},\quad \forall t\ge0\land\tau\in[0,{\pi}/{2\omega_D}]
\end{split}
\end{equation}
then we are told that the best measurement for QD is $\sigma_x$.


\begin{thebibliography}{10}
\bibitem{footnote_TLTS}
    We also calculated the relaxation times and their respective amplitudes by applying the procedure from ref.\cite{Keller_88} to the case of gypsum. The calculation agrees with experimental values from \cite{Jeene_AdvMR68}. A comparison between different strategies for calculating relaxation times using lattice sums will be published elsewhere.
\end{thebibliography}

\end{document}